\documentclass[twocolumn]{aastex631}
\usepackage{comment}

\usepackage{xcolor}

\shorttitle{Stellar Evolution}
\shortauthors{Sidhu et al.}

\graphicspath{{./}{figures/}}

\begin{document}

\title{The Interplay of Stellar Evolution and Collisions in Galactic Nuclei}

\correspondingauthor{Jaden S. Sidhu}
\email{jadensidhu2026@u.northwestern.edu}

\author{Jaden S. Sidhu}
\affiliation{Center for Interdisciplinary Exploration and Research in Astrophysics (CIERA), Northwestern University, 1800 Sherman Ave., Evanston, IL 60201, USA}

\author{Sanaea C. Rose}
\affiliation{Center for Interdisciplinary Exploration and Research in Astrophysics (CIERA), Northwestern University, 1800 Sherman Ave., Evanston, IL 60201, USA}

\author{Frederic A. Rasio}
\affiliation{Center for Interdisciplinary Exploration and Research in Astrophysics (CIERA), Northwestern University, 1800 Sherman Ave., Evanston, IL 60201, USA}

\begin{abstract}
Nuclear star clusters (NSCs) surrounding supermassive black holes (SMBHs) are among the densest stellar environments in the universe. In these environments, collisions can shape the stellar mass function and produce exotic stellar populations. In this work, we investigate how stellar collisions couple with stellar evolution 
in the inner parsec of an NSC. We simulate the evolution of a sample of $1000$ $1$ $M_\odot$ stars embedded in a uniform cluster of dynamically relaxed 0.5 $M_\odot$ stars.
Using COSMIC to evolve stellar properties in time
, we track the mass, radius, and evolutionary state of the stars as they collide in the cluster. Our results show that most stars within $0.1$~pc of the SMBH experience a collision while on the main-sequence. However, outside of this distance, stars collide during the red giant phase, when the stellar radius
increases dramatically. We find that the most common type of collision -- main-sequence or red giant -- over the lifetime of the cluster depends on the steepness of the stellar cusp, which determines the spatial distribution of the stars in the cluster.
These results show that stellar evolution plays a fundamental role in shaping the collisional history of stars in nuclear star clusters.
Lastly, we consider whether the closest known stars to the Milky Way's SMBH have experienced a collision. We estimate that several of the S-stars have a high probability of experiencing a collision over their main-sequence lifetime, perhaps with implications for their observed youth and properties.
\end{abstract}

\section{Introduction}
Like the Milky Way, many galaxies are believed to host a supermassive black hole (SMBH) \citep[e.g.,][]{Genzel+03,Kormendy04,FerrareseFord05,Ghez+05,KormendyHo13}. Surrounding the SMBH is a nuclear star cluster (NSC), a compact and densely-populated region of stars and stellar remnant \citep[e.g.,][]{Schodel+03,Ghez+05,Ghez+08,Gillessen+09,Gillessen+17}. The combination of a high density and large velocity dispersion leads to a greater probability of direct collisions \citep[e.g.,][]{BaileyDavies99,Dale+09,DaleDavies,Davies+11,Mastrobuono-B,Rose+20,Rose+22}. Previous work has shown that $\sim 10 \%$ of the stars experience at least one collision while on the main-sequence \citep[e.g.,][]{Rose+23}. Investigating the frequency and outcomes of these direct collisions is crucial in understanding the evolution and properties of nuclear star clusters.

Stellar collisions play a crucial role in shaping the evolutionary pathway of stars in the NSC, where close encounters and collisions are most probable. These interactions can produce atypical stellar remnants, inject gas into the environment, and modify the observed stellar population, potentially affecting interpretations of the NSC luminosity, star formation history, and mass function \citep[e.g.,][]{Freitag+02,Freitag+06,RubinLoeb,Mastrobuono-B,Rose+23,RoseMacLeod24}. Collisions and mergers of two main-sequence stars can produce rejuvenated objects, stars that have gained mass and have an extended evolution. The best known examples are {\em blue stragglers\/}, stars that appear to be younger, hotter, and more massive than all surrounding stars in a cluster \citep[e.g.,][]{Sills+97,Sills+01}. Mergers can rejuvenate the star by adding fresh hydrogen into the core, extending its main sequence lifetime. The presence of many merger products can bias age and population estimates of the host system \citep[e.g.,][]{Sills+10,MarinFranch+09}. Repeated collisions can trigger runaway growth in some environments. While these runaways are not expected in old NSCs, they could occur in young star clusters, where mergers can build very massive stars capable of collapsing into an intermediate-mass black hole (IMBH)\citep[e.g.,][]{PortegiesZwart,Gurkan+06}.

While some collisions lead to rejuvenation, others produce disrupted or stripped stars \citep[e.g.,][]{Balberg+13,Balberg24,Ryu+24b}. During grazing encounters, these stars lose a fraction of their mass as the impact's kinetic energy expels material from the outer layers of the star, leaving both ejecta in the surrounding environment and a lower mass remnant \citep[e.g.,][]{Davies+98,FreitagBenz,Freitag+08,Lai+93,RubinLoeb,Rauch99,Rose+23,RoseMacLeod24,Gibson+24,Rose+25}. The formation of both merger products and stripped stars can alter the mass function of the cluster \citep[][Parmerlee et al. in prep]{Rose+23}. 
Furthermore, collisions involving compact remnants, in our case a white dwarf-main-sequence interaction, can produce a wide range of outcomes. A head-on collision may generate a thermonuclear explosion, and if the combined mass reaches the Chandrasekhar limit, some simulations show that it may trigger a Type 1a-like reaction \citep[e.g.,][]{rosswog_tidal_2009,Raskin+09,Kushnir+13}. 



Previous work has explored the evolution and implications of stars in dense nuclear star clusters \citep[e.g.,][]{Rauch99,FreitagBenz,Freitag+06,Dale+09,Guillochon+09,macleod_spoon-feeding_2013,Merritt2013,AharonPerets16,Rose+23,RoseMacLeod24,Rose+25}. In this study, we build on these efforts by coupling stellar evolution with collisions. We begin with a  1$M_\odot$ main-sequence star and evolve it forward in time while allowing for collisions that modify its mass and radius. After each collision, the remnant’s updated properties serve as the initial conditions for continued evolution, enabling us to track how repeated mergers and stripping events alter the stellar population over time. This paper is organized as follows:

In Section~\ref{sec:methodology} we describe our overall methodology. Section~\ref{Dynamics} outlines the dynamical environment and the relevant physical processes involved in the NSC. Section~\ref{collisions} details the collisional physics, and Section~\ref{setup} provides a full description of the setup of our simulation setup. Section~\ref{sec:results} presents our results, focusing on the fraction of the stellar population that collides in Section~\ref{Fraction}, the collision rates by type in Section~\ref{Rates}, and trends in the collision properties in Section~\ref{Trends}. In Section~\ref{sec:S-stars} we discuss the implications of these results for the S-star population in the Galactic Center, including estimated collisional probabilities and their effects on the observed properties. Finally, we conclude in Section~\ref{sec:Conclusion}.

\section{Methods} \label{sec:methodology}


\subsection{Numerical Framework and Simulation Setup} \label{setup}

We follow a sample of evolving stars embedded in a fixed, uniform cluster. The sample population is initially distributed between 0.001 and 1 pc of the SMBH. All stars in the sample population have a mass of 1.0 $M_{\odot}$, which remains constant throughout the simulation. The orbital eccentricities are drawn from a thermal distribution, and the cluster is assumed to be spherically symmetric. 
We begin by sampling the initial semimajor axes of the evolving set of 1000 stars so that they are distributed uniformly in logarithmic distance between $0.001$ and $1$ pc. This set of initial conditions allows us to build a comprehensive picture of collision outcomes versus distance from the SMBH. However, later we also sample their initial semimajor axes to match the overall spatial distribution of the stars in the cluster, which reside on a cusp, described in more detail below.  

All stars in the background cluster are assumed to be $0.5$ M$_\odot$. We consider two possibilities for the spatial distribution of the stars in the fixed background cluster, both of which assume that the stars are dynamically relaxed. The underlying density profile of the cluster follows a power law as a function of $r_\bullet$, or distance from the SMBH,
\begin{equation}
    \rho(r_\bullet) = \rho_0\Bigg(\frac{r_\bullet}{r_0}\Bigg)^{-\alpha}
\end{equation}
where $\alpha$ sets the slope of the profile. The normalization is given by $\rho_0 = 1.35  \times10^6 M_\odot$ pc$^{-3}$ at $r_0 = 0.25$ pc \citep[][]{Genzel+10}. Theoretical expectations suggest $\alpha \approx 1.25-1.75$, and we test both extremes to compare their effects on the resulting collisional behavior \citep[][]{BahcallWolf76,Bar-Or+13,AlexanderHopman+09,Keshet+09,AharonPerets16,Buchholz+09,Do+09,Bartko+10,Gallego-Cano+18,Gallego+20,Schodel+14,Schodel+18,Schodel+20,LinialSari22}. The velocity dispersion of the cluster is approximated by the relation and the slope of the density profile $\alpha$:
\begin{equation}
    \sigma(r_\bullet) = \sqrt{\frac{GM_\bullet}{r_\bullet(1+\alpha)}},
        \label{eq.velocity}
\end{equation}
where $M_{\bullet}$ is the mass of the central SMBH, $r_\bullet$ is the distance from the Galactic Nucleus, and $\alpha$ is the slope of the density profile. Both the velocity dispersion and the stellar number density are important factors as they determine the frequency of collisions and interactions between a star and other compact objects \citep{Alexander99,AlexanderPfuhl14}.

\subsection{Stellar Evolution}

A key difference between this simulation and previous studies is that we explicitly include stellar evolution. We used COSMIC to model stellar evolution \citep[]{Breivik+20}. COSMIC (Compact Object Synthesis and Monte Carlo Investigation Code) is a rapid binary population synthesis code primarily used to model the evolution of binary stars, especially those that form compact objects. In our model, we used COSMIC to evolve a single star over time, using the default metallicity (0.02) assumed by the code. We evolved a grid of stars with masses ranging from $0.08$ to $30$ $M_\odot$ and created a data frame that recorded each star's mass ($M_\odot$), age (Myr), and radius ($R_\odot$). During the simulation, this data frame was interpolated to determine the radius and age corresponding to a given stellar mass.

\subsection{Two-Body Relaxation} \label{Dynamics}

A variety of dynamical processes drive the evolution of a nuclear star cluster. These dense systems are typically dominated by the gravitational potential of an SMBH \citep[e.g.,][]{Genzel+03,Schodel+03,Gillessen+17}. Stars also experience gravitational perturbations from other stars in the cluster. Over time, these small perturbations accumulate and cause a significant change in a star's orbital energy and angular momentum, a process known as two-body relaxation \citep[e.g.,][]{Spitzer1987,BinneyTremaine,AlexanderHopman+09}.

The relaxation time, $t_r$, is given by
\begin{equation}
    t_r = 0.34\frac{\sigma^3}{G^2M_*^2n\ln\Lambda},
\end{equation}
where $\sigma$ is the velocity dispersion, $G$ is the gravitational constant, $M_*$ is the mass of the star, $n$ is the stellar number density, and $\ln\Lambda$ is the Coulomb logarithm. In an NSC hosting an SMBH, the relaxation time is approximately $10^8-10^{10}$ years \citep[e.g.,][]{Merritt2013,Bar-Or+13}. In a dense environment, the relaxation time falls below the age of the universe, making the cluster a collisional stellar system \citep[e.g.,][]{HopmanAlexander05,BinneyTremaine,Freitag+06,Dale+09}. In our simulations, we take a Monte Carlo approach and follow the evolution of $1000$ $1$~M$_\odot$ stars with initial orbits as described in Section~\ref{setup}. We account for the effects of relaxation by allowing the orbital parameters of the stars to undergo a random walk over the characteristic relaxation timescale (see \citet{Rose+22} for the implementation of this process in the code and \citet{Naoz+22} for the full equations).



\subsection{Direct Collisions} \label{collisions}

The high stellar densities and velocity dispersions that are present near the SMBH lead to a high rate of direct stellar collisions \citep[e.g.,][]{DuncanShapiro83,DaleDavies,Antonini+11}. The collision timescale for a star can be determined by:
\begin{equation}
t_{coll} = \frac{1}{n(r) \sigma_{coll}v(r)},
\end{equation}
where $n$ is the stellar number density, $\sigma_{coll}$ is the collision cross section, and $v$ is the relative velocity, taken to be the velocity dispersion. For the region we are studying, within $\sim 1$ pc of the SMBH, the typical stellar number density is about $10^8 - 10^{10}$ pc$^{-3}$ \citep[e.g.,][]{Genzel+10,Merritt2013}. The collisional cross section is defined as
\begin{equation}
    \sigma_{coll} = \pi r_{coll}^2 \Big(1+\frac{v^2_{esc}}{v_{\infty}^2} \Big),
\end{equation}
where $r_{coll}$ is the sum of the stellar radii, $v_\infty$ is the relative velocity at infinity (before gravitational focusing), and $v_{esc}$ is the escape velocity,
\begin{equation}
    v_{esc} = \sqrt{2G\frac{(M_1+M_2)}{r_{coll}}},
\end{equation}
where $G$ is the gravitational constant, and $M_1$ and $M_2$ are the masses of the two objects.

The outcome of the collision depends on the impact parameter, mass ratio, and relative velocity \citep[e.g.,][]{Lai+93}. Previous studies have used smooth-particle hydrodynamics (SPH) to provide detailed insight into the effect of an individual collision on the star \citep[e.g.,][]{Lai+93,Lombardi+96,Lombardi+02,Sills+01,FreitagBenz}. Our simulation incorporates fitting formulae from previous SPH simulations to predict the mass loss following a collision and determine whether it results in a merger (see \citet{Rauch99} for the fitting formulae and \citet{Rose+23} for the implementation in the code). This frameworks allows us to probabilistically sample collision events near the SMBH. When a collision occurs, stellar properties such as mass and radius are updated, and the post-collision object is then evolved forward in time using COSMIC until the next collision. 

\section{Stopping Conditions}

The runtime of the simulation is 10 Gyr. However, we also included several stopping conditions related to the evolution of stars in our sample. At each timestep, the star's main-sequence lifetime, $t_{MS}$, is calculated from its mass using
\begin{equation}
    t_{MS} = 10^{10}\Big(\frac{M}{M_\odot}\Big)^{-2.5}.
\end{equation}
The simulation is terminated when the star's age exceeds this lifetime. We also terminate the simulation if significant mass loss from the collisions causes the mass of a star to fall below 0.08 $M_\odot$. 
Furthermore, two-body relaxation can alter the orbits of the sample stars, as can stellar collisions \citep{RoseMockler25}. These interactions can cause stars to pass within the SMBH's tidal radius and become ruptured, producing a tidal disruption event (TDE) \citep[e.g.,][]{Hills1975,Rees1988,Guillochon+13,Stone+17}. 
The simulation terminates a star if it is tidally disrupted by the SMBH. Furthermore, occasionally a collision ejects a star from the cluster \citep[][]{RoseMockler25}. If the final orbit of a sample star is determined to be unbound from the SMBH, we record the event and terminate the star's evolution within the cluster.

\section{Results} \label{sec:results}

\begin{figure}
    \centering
    \includegraphics[width=\columnwidth]{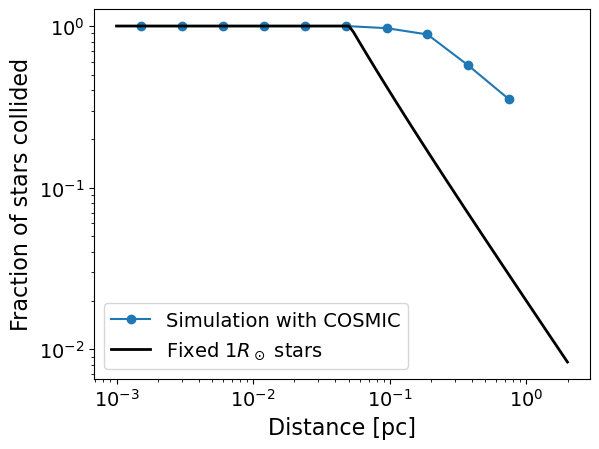}
    \caption{Fraction of stars collided vs distance in pc. The black curve is the analytical prediction for the collision timescale $t_{coll}$. The blue curve is the simulated fraction of stars that have undergone a collision. The simulation focused on a sample of  1000 1 $M_{\odot}$ stars evolving over time in cluster of 0.5 $M_{\odot}$ stars. All stars are expected to collide close within $0.1$~pc the SMBH, and the analytical prediction and simulation are in agreement. However, outside this distance, we have more collisions in the simulation than expected in the analytical prediction based on the main-sequence properties of the star. The stars in our simulation evolve over time, eventually becoming red giants. The red giant stars have a larger cross section, which leads to more collisions compared to the analytical prediction, which assumes the 1 $R_{\odot}$ stars do not evolve or change. Stellar evolution can therefore significantly enhance the fraction of the population that experiences a direct collision. }
    \label{fig:fractioncollided}
\end{figure}

\subsection{Fraction of Stars Experiencing Collisions} \label{Fraction}

We begin by comparing analytic expectations for the fraction of stars that experience a collision based on their main-sequence radius versus evolving stellar radii. Figure~\ref{fig:fractioncollided} compares the fraction of stars that have undergone a collision in the simulation with stellar evolution (blue line) to the analytical prediction using the collision timescale for main-sequence stars (black line) as a function of distance from the SMBH. The collision timescale, $t_{coll}$, is given by Equation 4 and was computed using the local stellar density, velocity dispersion, and stellar radius assuming a $1M_\odot$ and $1R_\odot$ star in a background of $0.5M_\odot$ stars. The density profile and velocity dispersion were computed for $\alpha = 1.75$ and a SMBH mass of $M_{BH} = 4 \times10^6M_\odot$. Given the timescale and simulation duration of 10 Gyr, the fraction of the population expected to collide as a function of distance from the SMBH can be estimated as the maximum between $\frac{1 \times 10^{10} \, \mathrm{yr}}{t_{coll}}$ and $1$. 

Within $0.1$~pc, the analytical prediction for main-sequence stars and the simulation results are in agreement. The high density profile and velocity dispersion result in a collision timescale that is less than the main-sequence lifetime of the stars \citep[see also][]{RoseMacLeod24}, making collisions highly likely in this region while the stars are still on the main-sequence. 1 $M_\odot$ stars in this region should collide at least once before evolving into red giants. At larger radii, however, the simulation produced more collisions compared to the black curve. This discrepancy arises because the analytical prediction assumes a fixed stellar radii of $1$ $R_\odot$, whereas the simulation  includes stellar evolution. As stars evolve off of the main-sequence and into red giants, their radii increase, which also increases their cross section. This significantly raises their likelihood of colliding with other stars relative to the non-evolving main-sequence star in the analytical prediction. From our simulation with COSMIC, we estimate about 87\% of the stars residing within 1 pc of the SMBH experience a collision. These results highlight the importance of modeling stellar evolution, as it can significantly alter the collision prospects of stars in dense environments. 

\subsection{Collision Trends with Distance} \label{Trends}

\begin{figure*}
    
    \centering
    \includegraphics[width=\linewidth]{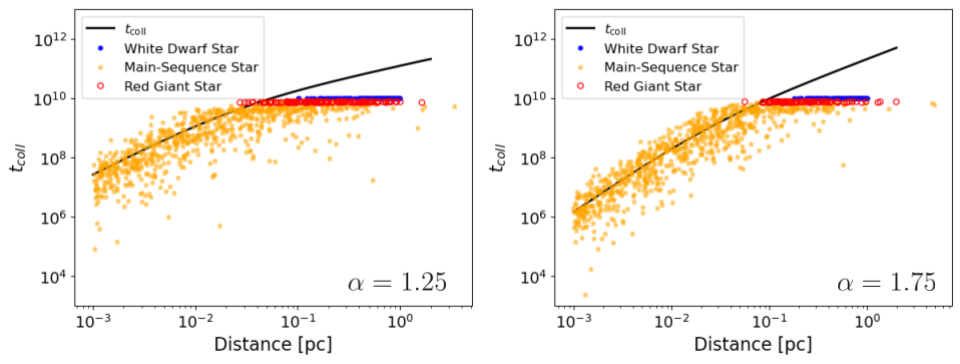}
    \caption{The time of a star's first collision in years vs. the semimajor axis of its orbit about the SMBH in pc. The background is composed of 0.5 $M_{\odot}$ stars evenly spaced between 1 pc of the galactic center. We consider 1000 1 $M_{\odot}$ stars distributed uniformly in log distance from the SMBH, to understand the distance trends of the collisions. We test two values for the slope of the background cluster density profile. The three symbols represent the evolutionary stage of the star when the collision occurs: main-sequence (orange stars), red giant (red open circles), or white dwarf (blue circle).The black curve is the analytical prediction for $t_{coll}$ for main-sequence stars that do not evolve.} 
    \label{fig:distancetrends}
\end{figure*}

To complement Figure~\ref{fig:fractioncollided}, we investigate trends in the time of collisions versus distance from the SMBH. In Figure~\ref{fig:distancetrends}, we plot the time of the first collision for each sample star versus its semimajor axis. Symbols represent the evolutionary stage of the star at that time. We note that the $0.5$ M$_\odot$ background stars remain on the main-sequence for the entirety of the simulation, so the categories of collision (main-sequence, red giant, or white dwarf) denote the evolutionary stage of the sample star we are following. We also plot the analytical prediction (black line) for collisions between main-sequence stars. The simulations presented in the Figure sampled stars uniformly in log distance. We show simulations for two density profiles of the background cluster: $\alpha = 1.25$ and $\alpha = 1.75$. In the simulation with the Bahcall-Wolf profile ($\alpha = 1.75$, \citet{BahcallWolf76}), the main-sequence collision timescale is shorter and the denser inner region leads to collisions occurring at earlier times. The shorter collision timescale for $\alpha = 1.75$ also means that there are overall more main-sequence collisions and that this collision dominates out to larger distances from the SMBH. The shallower cusp $\alpha = 1.25$, with a relatively longer collision timescale, means that more stars avoid colliding during the main-sequence phase, but do experience their first collision as red giants. For both cusps, red giant collisions dominate closer to the edge of the sphere of influence ($\sim 1$ pc). Red giant stars have a high collision probability due to their larger radii and cross sections \citep[see also, e.g.,][]{BaileyDavies99,Dale+09,Ryu+24b}.
The simulation is limited to $10^{10}$ years, which sets a cutoff in the figure. While we plot only the time of the first collision and use symbols representative of the star's evolutionary phase at that time, we note that stars that experience one or more collisions on the main-sequence can also experience another collision after evolving into a red giant.

\subsection{Rates of Collision by Type} \label{Rates}

\begin{figure*}
    \centering
    \includegraphics[width=\linewidth]{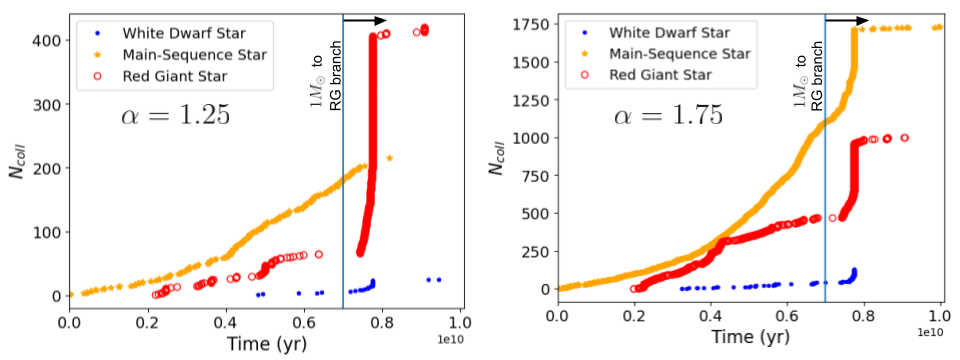}
    \caption{Cumulative number of collisions $N_{coll}$ vs. time in yr for 1000 1 $M_{\odot}$ star evolving over time in a dense background cluster of 0.5 $M_{\odot}$ stars residing on a cusp with slope $\alpha$. The three symbols define the star when the collision occurs: main-sequence (orange stars), red giant (red open circles), or white dwarf (blue circle). The sudden spike in collisions in the red giant phase coincides with $1$~M$_\odot$ stars evolving into red giants. Most of the stars live around the 1 pc distance from the galactic center, so there are not as many collisions until the star evolves into a red giant.}
    \label{fig:n_collisions}
\end{figure*}

To understand how different stellar populations contribute to the overall collision count, we next evaluated the collision rates at each evolutionary stage. From hereon, all simulations shown follow a set of 1000 stars whose semimajor axes are sampled such that they share the same spatial distribution of the background cluster and reside on a cusp. Sampling the semimajor axes of the stars in this way ensures that the relative rates of each type of collision are proportionate to their overall prevalence in the cluster. For example, if $200$ main-sequence collisions occur for our sample of $1000$ stars over 10 Gyr, we can expect a total of roughly $8 \times 10^5$ main-sequence collisions to occur for the $\sim 4 \times 10^6$ stars which reside in this region ($\lesssim 1$ pc of the SMBH). Figure~\ref{fig:distancetrends} shows the cumulative number of collisions, $N_{coll}$, over time for a $1 M_\odot$ star evolving in a dense stellar cusp of dynamically relaxed $0.5 M_\odot$ stars repeated 1000 times. These simulations assume density profiles with slope $\alpha = 1.25$ (left plot) or $\alpha = 1.75$ (right plot). Each symbol represents the evolutionary phase of the star at the time of collision: white dwarf (blue points), main-sequence (orange stars), and red giant (red open circles). The slope of each curve represents the rate of collisions. 

The spike in the red giant collisions in Figure~\ref{fig:n_collisions} corresponds to when $1$ M$_\odot$ stars ascend the red giant branch. We have added a vertical line to the plot to mark the onset of the red giant phase. Their vastly expanded cross-section produces a multitude of collisions all within the red giant's short lifespan. Red giant collisions before the vertical line come from stars that had gained mass from an earlier collisions. This additional mass moves the star to a higher-mass evolutionary track, increasing the hydrogen-burning rate and accelerating its evolution. Once the red giant phase has passed, the remaining white dwarf has a smaller radius and cross section, which leads to very few collisions.

For our sample of 1000 stars with $\alpha = 1.25$, linear fits for the main-sequence curve gives us a collision rate of $3.1\times10^{-8}$ per year, or 31.0 collisions per Gyr. Linear fits for the red giant curve between the time period of $7\times10^9$ and $8\times10^9$ years yield collision rate of $9.1\times10^{-7}$ per year, or 911.9 collisions per Gyr. Since the number of white dwarf-main-sequence star collisions are low over the course of the simulation (total of 25 over 10 Gyr), we do not perform a linear fit to determine the rate. Rather, we simply estimate the rate by scaling to the number of stars expected in the sphere of influence and dividing by $10$ Gyr. Within our sample, we have roughly $2.5\times10^{-9}$ white dwarf-main-sequence star collisions per year, or 2.5 collisions per Gyr. If we estimate these rates for the entire cluster, we get $1.2\times10^{5}$ collisions per Gyr for main-sequence stars, $3.6\times10^{6}$ collisions per Gyr for red giants, and $1.0\times10^{4}$ collisions per Gyr for white dwarfs.

Similarly for our sample of 1000 stars with $\alpha = 1.75$, the collision rate of main-sequence stars, red giants, and white dwarfs are $1.8\times10^{-7}$, $1.2\times10^{-6}$, and $1.2\times10^{-8}$ collisions per year, or 178.8, 1163.2, 12.7 collisions per Gyr, respectively. The red giant collision rate during its peak is between 6 to 29 times higher than the main-sequence collision rate. Estimating the rates for the entire cluster, we get $7.2\times10^{5}$ collisions per Gyr for main-sequence stars, $4.7\times10^{6}$ collisions per Gyr for red giants, and $5.1\times10^{4}$ collisions per Gyr for white dwarfs. We can compare this result with the number of collisions seen in \citet{FreitagBenz_confproceedings_2002}. While they do not include stellar evolution, they estimate about $4\times10^{5}$ collisions per Gyr for a main-sequence-main-sequence collision, which falls in the same order of magnitude as our results. Altogether, we find that the most common type of collision from our simulations, be that main-sequence-main-sequence or red giant-main-sequence, depends on the assumed stellar cusp. A cusp with $\alpha = 1.75$ leads to more main-sequence collisions because the collision timescale is shorter and fewer stars avoid colliding before becoming red giants, as discussed in the previous section. The shorter timescale also means that stars in the dense inner $0.1$~pc can also experience multiple collisions while on the main-sequence.



\begin{figure*}
    \centering
    \includegraphics[width=\linewidth]{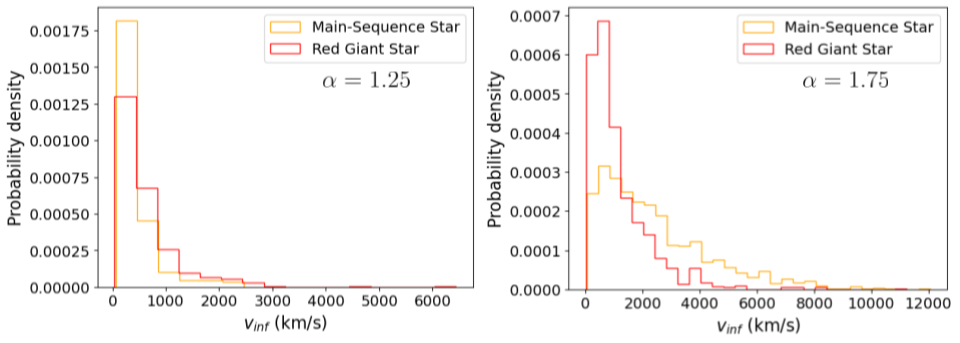}
    \caption{Distribution of the relative velocity at infinity ( $v_\infty$) for stellar collisions involving white dwarfs (blue), main-sequence stars (orange), and red giants (red). The red giants experience a broader distribution at higher velocities, while the main-sequence and white dwarf stars are concentrated at lower values, which is caused by the different orbital energies and interaction dynamics near the SMBH.}
    \label{fig:v_inf}
\end{figure*}

Next, we examine kinematic trends by analyzing the distribution of incoming velocities for these collisions. Figure~\ref{fig:v_inf} depicts the distribution of relative velocities at infinity ($v_{\infty}$) for stellar collisions with a density profile value of $\alpha = 1.25$ and $\alpha = 1.75$. The nuclear star cluster is dominated by the gravitational potential of the SMBH.
The distance of the star from the SMBH determines the relative velocity (see Eq.~\ref{eq.velocity}). Since both main-sequence stars and red giant stars occupy similar orbital radii, they experience essentially the same gravitational potential, resulting in nearly identical velocity dispersions, particularly for $\alpha = 1.25$. For $\alpha = 1.75$, the main-sequence stellar collisions have a more substantial high-velocity tail compared to the red giants as the red giant collisions do not extend to as close in regions to the SMBH. Furthermore, the relative speeds of the collisions tend to be generally lower for $\alpha = 1.25$ due to the weak dependence of the velocity dispersion on $\alpha$. 



\begin{figure*}
    \centering
    \includegraphics[width=\linewidth]{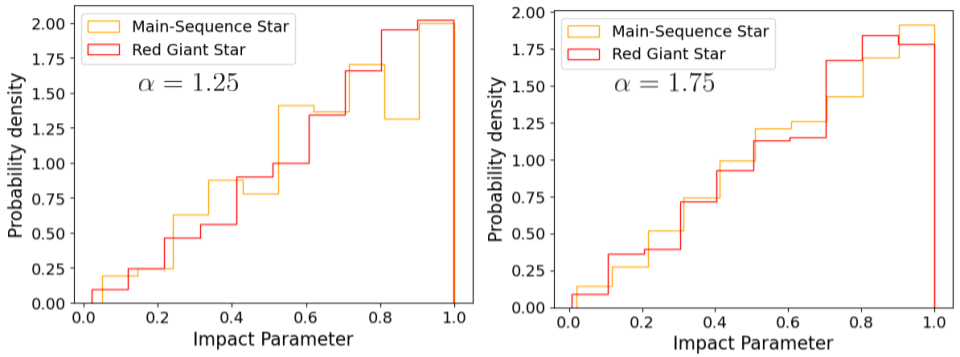}
    \caption{Distribution of impact parameters for stellar collisions involving white dwarfs (blue), main-sequence stars (orange), and red giants (red). The histogram shows that the collisions tend to occur with larger impact parameters.}
    \label{fig:b}
\end{figure*}

Finally, Figure~\ref{fig:b} shows the distribution of impact parameters for stellar collisions for density profiles $\alpha = 1.25$ and $\alpha = 1.75$. Smaller impact parameters correspond to head-on, direct collision, whereas larger values correspond to grazing, near fly-by encounters. In the high-velocity environment of the nuclear star cluster, the collision cross-section (Equations 5 and 6) is set primarily by the stellar velocity dispersion. The velocity dispersion within the cluster spans hundreds to thousands of km/s due to the steep gravitational potential of the SMBH, and as a result, gravitational focusing is weak.
As shown in Figure~\ref{fig:v_inf}, both main-sequence and red giant star collisions share similar relative velocities due to their common dynamical environment. For the same reason, the collisions have similar distributions of impact parameters, with larger impact parameters, or grazing collisions, being more likely. 

\section{Implications for S-stars} \label{sec:S-stars}

\begin{figure*}
    \centering
    \includegraphics[width=\linewidth]{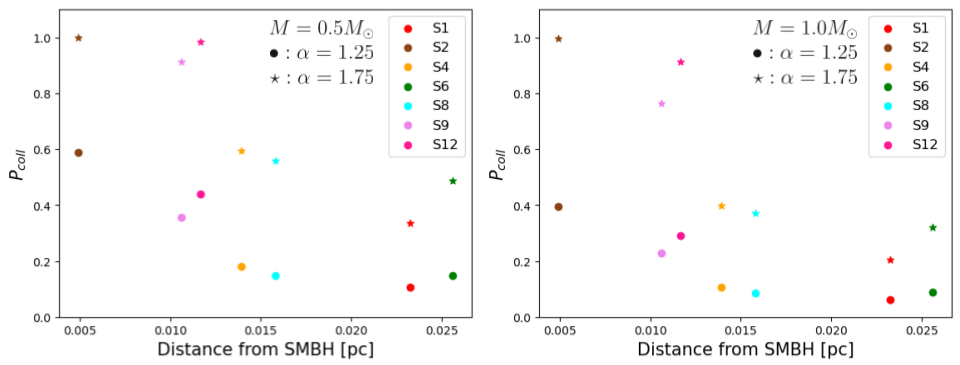}
    \caption{Probability of at least one collision to occur vs. the distance from the SMBH in pc for the s-stars. Results are shown for two different background populations: one composed entirely of $0.5M_\odot$ stars (left), consistent with our simulation set-up and one composed entirely of $1.0M_\odot$ stars (right). The marker shape shows the two extremes of the stellar density profile, with the filled circle representing $\alpha = 1.25$ and the star symbols representing $\alpha = 1.75$.}
    \label{fig:s_stars}
\end{figure*}

The S-stars are young, massive stars that have been observed on isotropically-distributed close-in orbits around the SMBH in the Milky Way's Galactic center \citep[e.g.,][]{Ghez+03,Ghez+08,Gillessen+09,Genzel+10}. The origins of these stars are uncertain and are key to understanding star formation and stellar dynamics in extreme gravitational environments \citep[][]{Levin+03,Alexander05,Genzel+03}. Given the prevalence of main-sequence collisions in the inner $0.1$~pc, we consider whether the S-stars may have been affected by a previous collision \citep[e.g.,][]{FreitagBenz,Dale+09}. The general likelihood that some of the S-stars have been impacted by a collision has been noted in the past \citep[][]{Freitag+08_confproceeding}, and sequential collisions have also been proposed as a potential origin for the S-stars \citep[][]{Rose+23,RoseMockler25}. In this study, we estimate the probability of collision for 7 S-stars over their main-sequence lifetimes based on their current masses and orbits. We obtain the observed orbital parameters and masses for the S-stars from \citet{Habibi+17} and \citet{Gillessen+17}. We calculate the collision timescale of each S-star and its expected main-sequence lifetime and estimate the probability of collision; this is shown in Figure~\ref{fig:s_stars}. We find that several S-stars have collision probabilities above 50\% for the maximum $\alpha = 1.75$ value, while the minimum value of $\alpha = 1.25$ still gives collision probabilities over 20\% for some of the S-stars. We note that observations of the present-day Galactic center suggest a shallower cusp, closer to our $\alpha = 1.25$ \citep[e.g.,][]{Gallego-Cano+18,Schodel+18}. It is also important to note that using a density profile normalized by the M-sigma relation based on \citet{Tremaine+02} yields collision probabilities that are a factor of two higher than those in the figure, but we have chosen to show the more conservative prediction.

This implies that collisions in our Galactic nucleus have likely impacted at least some of the observed S-stars. If a star has collided in the past, then its current properties, including luminosity, apparent age, temperature, and rotational velocity, do not accurately reflect isolated stellar evolution \citep[e.g.,][]{Lombardi+02,Glebbeek+13,RoseMacLeod24}. Instead the S-stars may be the remnant of a merger, like a rejuvenated or blue straggler star \citep[e.g.,][]{Sills+97,Sills+01,Glebbeek+13}
A previous merger may provide a natural explanation for why the S-stars appear to be young and also suggests that the inferred age of the S-star population may be underestimated. Future work will explore the implications of a previous collision for the S-stars. 


\section{Conclusions} \label{sec:Conclusion}


In this work, we investigated how direct stellar collisions shape the evolution of stars in dense nuclear star clusters. Our goal was to quantify when and in what evolutionary stage stars are most likely to experience collisions and explore the results of these interactions for the stellar population. We follow the evolution of a sample of 1000 $1 M_\odot$ stars orbiting a $4 \times 10^6$ M$_\odot$ SMBH in a dynamical relaxed background cluster of $0.5 M_\odot$ stars. We use a code developed by \citet{Rose+22,Rose+23} and \citet{RoseMockler25}, which accounts for the effects two-body relaxation and direct collisions using a statistical approach. Consistent with those previous studies, we use fitting formulae \citep{Rauch99} from SPH simulations to predict the precise outcome of a collision, such as the mass loss and whether or not a merger occurs. We build upon previous work by using COSMIC \citep[]{Breivik+20} to include a treatment for stellar evolution. Including stellar evolution is essential; as stars evolve off the main sequence and expand, their collision cross-sections grow substantially, altering both the likelihood and timing of direct collisions.

From our simulations with stellar evolution, we find that $\sim 85 \%$ of $1$ M$_\odot$ stars within 1 pc of the SMBH will experience a collision at some point during their evolution, before forming a compact object. This percentage is higher than the $\sim 10$ to $40 \%$ predicted by simulations without stellar evolution \citep[][Parmerlee et al. in prep.]{Rose+23}. Stars within $0.1$ pc experience their first collision on the main-sequence, however whether most collisions occur during the main-sequence or red giant phase depends strongly on the slope of the density profile. In a steeper cusp ($\alpha = 1.75$), high densities drive many main-sequence collisions. On the other hand, in a shallower cusp ($\alpha = 1.25$), more stars avoid early collisions and experience their first collision as red giants, when their expanded radii greatly enhance the cross-section. White dwarfs experience little to no collisions after forming because of their small radii. The white dwarf-main-sequence star collision rate is also limited by the fact that most of our stars only become white dwarfs in the last few Gyr of the simulation. Across all models, collisions during the red giant phase contribute significantly to the total collisions of the system and are largely responsible for boosting the overall fraction of stars that collide.

These collisions can produce unusual stellar products. In particular, collisions may rejuvenate stars by supplying fresh hydrogen to their cores, extending their lifetime and creating blue straggler-like objects. Using our collision rates, we estimate the probability that seven well-studied S-stars on close-in orbits about the Milky Way's SMBH have experienced a collision during their main-sequence lifetime. Our results suggest that the extreme dynamical environment near the SMBH makes collisions a realistic possibility for stars on these orbits.

In summary, the collision history of a star in a NSC is strongly dependent on its orbital radius, the local density profile, and its evolutionary stage. Future work will focus on a more detailed comparison between predicted collision products and the observed properties of the S-star population. Furthermore, future work will extend this model to account for an initial mass function of the stars. Our findings highlight the need for  simulations to explore the rich interplay between hydrodynamics and stellar evolution for the full spectrum of collisions expected in NSCs. 

\begin{acknowledgments}

SCR thanks the Lindheimer Postdoctoral Fellowship for support. This research was supported in part through the computational resources and staff contributions provided for the Quest high-performance computing facility at Northwestern University, which is jointly supported by the Office of the Provost, the Office for Research, and Northwestern University Information Technology. This work made use of the following software packages: \texttt{astropy} \citep{astropy:2013,astropy:2018,astropy:2022}, \texttt{matplotlib} \citep{Hunter:2007}, \texttt{numpy} \citep{numpy}, \texttt{pandas} \citep{mckinney-proc-scipy-2010,pandas_17806077}, \texttt{python} \citep{python}, \texttt{scipy} \citep{2020SciPy-NMeth,scipy_17467817}, \texttt{COSMIC} \citep{Breivik2020,COSMIC_15855958}, \texttt{Cython} \citep{cython:2011}, \texttt{h5py} \citep{collette_python_hdf5_2014,h5py_7560547}, \texttt{schwimmbad} \citep{schwimmbad}, \texttt{seaborn} \citep{Waskom2021}, and \texttt{tqdm} \citep{tqdm_14231923}. Software citation information aggregated using \texttt{\href{https://www.tomwagg.com/software-citation-station/}{The Software Citation Station}} \citep{software-citation-station-paper,software-citation-station-zenodo}.
\end{acknowledgments}

\bibliography{sample631}{}

\begin{thebibliography}{}
\expandafter\ifx\csname natexlab\endcsname\relax\def\natexlab#1{#1}\fi
\providecommand{\url}[1]{\href{#1}{#1}}
\providecommand{\dodoi}[1]{doi:~\href{http://doi.org/#1}{\nolinkurl{#1}}}
\providecommand{\doeprint}[1]{\href{http://ascl.net/#1}{\nolinkurl{http://ascl.net/#1}}}
\providecommand{\doarXiv}[1]{\href{https://arxiv.org/abs/#1}{\nolinkurl{https://arxiv.org/abs/#1}}}

\bibitem[{{Aharon} \& {Perets}(2016)}]{AharonPerets16}
{Aharon}, D., \& {Perets}, H.~B. 2016, \apjl, 830, L1, \dodoi{10.3847/2041-8205/830/1/L1}

\bibitem[{{Alexander}(1999)}]{Alexander99}
{Alexander}, T. 1999, \apj, 527, 835, \dodoi{10.1086/308129}

\bibitem[{{Alexander}(2005)}]{Alexander05}
---. 2005, \physrep, 419, 65, \dodoi{10.1016/j.physrep.2005.08.002}

\bibitem[{{Alexander} \& {Hopman}(2009)}]{AlexanderHopman+09}
{Alexander}, T., \& {Hopman}, C. 2009, \apj, 697, 1861, \dodoi{10.1088/0004-637X/697/2/1861}

\bibitem[{{Alexander} \& {Pfuhl}(2014)}]{AlexanderPfuhl14}
{Alexander}, T., \& {Pfuhl}, O. 2014, \apj, 780, 148, \dodoi{10.1088/0004-637X/780/2/148}

\bibitem[{{Antonini} {et~al.}(2011){Antonini}, {Lombardi}, \& {Merritt}}]{Antonini+11}
{Antonini}, F., {Lombardi}, James~C., J., \& {Merritt}, D. 2011, \apj, 731, 128, \dodoi{10.1088/0004-637X/731/2/128}

\bibitem[{{Astropy Collaboration} {et~al.}(2013){Astropy Collaboration}, {Robitaille}, {Tollerud}, {Greenfield}, {Droettboom}, {Bray}, {Aldcroft}, {Davis}, {Ginsburg}, {Price-Whelan}, {Kerzendorf}, {Conley}, {Crighton}, {Barbary}, {Muna}, {Ferguson}, {Grollier}, {Parikh}, {Nair}, {Unther}, {Deil}, {Woillez}, {Conseil}, {Kramer}, {Turner}, {Singer}, {Fox}, {Weaver}, {Zabalza}, {Edwards}, {Azalee Bostroem}, {Burke}, {Casey}, {Crawford}, {Dencheva}, {Ely}, {Jenness}, {Labrie}, {Lim}, {Pierfederici}, {Pontzen}, {Ptak}, {Refsdal}, {Servillat}, \& {Streicher}}]{astropy:2013}
{Astropy Collaboration}, {Robitaille}, T.~P., {Tollerud}, E.~J., {et~al.} 2013, \aap, 558, A33, \dodoi{10.1051/0004-6361/201322068}

\bibitem[{{Astropy Collaboration} {et~al.}(2018){Astropy Collaboration}, {Price-Whelan}, {Sip{\H{o}}cz}, {G{\"u}nther}, {Lim}, {Crawford}, {Conseil}, {Shupe}, {Craig}, {Dencheva}, {Ginsburg}, {Vand erPlas}, {Bradley}, {P{\'e}rez-Su{\'a}rez}, {de Val-Borro}, {Aldcroft}, {Cruz}, {Robitaille}, {Tollerud}, {Ardelean}, {Babej}, {Bach}, {Bachetti}, {Bakanov}, {Bamford}, {Barentsen}, {Barmby}, {Baumbach}, {Berry}, {Biscani}, {Boquien}, {Bostroem}, {Bouma}, {Brammer}, {Bray}, {Breytenbach}, {Buddelmeijer}, {Burke}, {Calderone}, {Cano Rodr{\'\i}guez}, {Cara}, {Cardoso}, {Cheedella}, {Copin}, {Corrales}, {Crichton}, {D'Avella}, {Deil}, {Depagne}, {Dietrich}, {Donath}, {Droettboom}, {Earl}, {Erben}, {Fabbro}, {Ferreira}, {Finethy}, {Fox}, {Garrison}, {Gibbons}, {Goldstein}, {Gommers}, {Greco}, {Greenfield}, {Groener}, {Grollier}, {Hagen}, {Hirst}, {Homeier}, {Horton}, {Hosseinzadeh}, {Hu}, {Hunkeler}, {Ivezi{\'c}}, {Jain}, {Jenness}, {Kanarek}, {Kendrew}, {Kern}, {Kerzendorf}, {Khvalko}, {King}, {Kirkby}, {Kulkarni},
  {Kumar}, {Lee}, {Lenz}, {Littlefair}, {Ma}, {Macleod}, {Mastropietro}, {McCully}, {Montagnac}, {Morris}, {Mueller}, {Mumford}, {Muna}, {Murphy}, {Nelson}, {Nguyen}, {Ninan}, {N{\"o}the}, {Ogaz}, {Oh}, {Parejko}, {Parley}, {Pascual}, {Patil}, {Patil}, {Plunkett}, {Prochaska}, {Rastogi}, {Reddy Janga}, {Sabater}, {Sakurikar}, {Seifert}, {Sherbert}, {Sherwood-Taylor}, {Shih}, {Sick}, {Silbiger}, {Singanamalla}, {Singer}, {Sladen}, {Sooley}, {Sornarajah}, {Streicher}, {Teuben}, {Thomas}, {Tremblay}, {Turner}, {Terr{\'o}n}, {van Kerkwijk}, {de la Vega}, {Watkins}, {Weaver}, {Whitmore}, {Woillez}, {Zabalza}, \& {Astropy Contributors}}]{astropy:2018}
{Astropy Collaboration}, {Price-Whelan}, A.~M., {Sip{\H{o}}cz}, B.~M., {et~al.} 2018, \aj, 156, 123, \dodoi{10.3847/1538-3881/aabc4f}

\bibitem[{{Astropy Collaboration} {et~al.}(2022){Astropy Collaboration}, {Price-Whelan}, {Lim}, {Earl}, {Starkman}, {Bradley}, {Shupe}, {Patil}, {Corrales}, {Brasseur}, {N{"o}the}, {Donath}, {Tollerud}, {Morris}, {Ginsburg}, {Vaher}, {Weaver}, {Tocknell}, {Jamieson}, {van Kerkwijk}, {Robitaille}, {Merry}, {Bachetti}, {G{"u}nther}, {Aldcroft}, {Alvarado-Montes}, {Archibald}, {B{'o}di}, {Bapat}, {Barentsen}, {Baz{'a}n}, {Biswas}, {Boquien}, {Burke}, {Cara}, {Cara}, {Conroy}, {Conseil}, {Craig}, {Cross}, {Cruz}, {D'Eugenio}, {Dencheva}, {Devillepoix}, {Dietrich}, {Eigenbrot}, {Erben}, {Ferreira}, {Foreman-Mackey}, {Fox}, {Freij}, {Garg}, {Geda}, {Glattly}, {Gondhalekar}, {Gordon}, {Grant}, {Greenfield}, {Groener}, {Guest}, {Gurovich}, {Handberg}, {Hart}, {Hatfield-Dodds}, {Homeier}, {Hosseinzadeh}, {Jenness}, {Jones}, {Joseph}, {Kalmbach}, {Karamehmetoglu}, {Ka{l}uszy{'n}ski}, {Kelley}, {Kern}, {Kerzendorf}, {Koch}, {Kulumani}, {Lee}, {Ly}, {Ma}, {MacBride}, {Maljaars}, {Muna}, {Murphy}, {Norman}, {O'Steen},
  {Oman}, {Pacifici}, {Pascual}, {Pascual-Granado}, {Patil}, {Perren}, {Pickering}, {Rastogi}, {Roulston}, {Ryan}, {Rykoff}, {Sabater}, {Sakurikar}, {Salgado}, {Sanghi}, {Saunders}, {Savchenko}, {Schwardt}, {Seifert-Eckert}, {Shih}, {Jain}, {Shukla}, {Sick}, {Simpson}, {Singanamalla}, {Singer}, {Singhal}, {Sinha}, {Sip{H{o}}cz}, {Spitler}, {Stansby}, {Streicher}, {{{S}}umak}, {Swinbank}, {Taranu}, {Tewary}, {Tremblay}, {Val-Borro}, {Van Kooten}, {Vasovi{'c}}, {Verma}, {de Miranda Cardoso}, {Williams}, {Wilson}, {Winkel}, {Wood-Vasey}, {Xue}, {Yoachim}, {Zhang}, {Zonca}, \& {Astropy Project Contributors}}]{astropy:2022}
{Astropy Collaboration}, {Price-Whelan}, A.~M., {Lim}, P.~L., {et~al.} 2022, \apj, 935, 167, \dodoi{10.3847/1538-4357/ac7c74}

\bibitem[{{Bahcall} \& {Wolf}(1976)}]{BahcallWolf76}
{Bahcall}, J.~N., \& {Wolf}, R.~A. 1976, \apj, 209, 214, \dodoi{10.1086/154711}

\bibitem[{{Bailey} \& {Davies}(1999)}]{BaileyDavies99}
{Bailey}, V.~C., \& {Davies}, M.~B. 1999, \mnras, 308, 257, \dodoi{10.1046/j.1365-8711.1999.02740.x}

\bibitem[{{Balberg}(2024)}]{Balberg24}
{Balberg}, S. 2024, \apj, 962, 150, \dodoi{10.3847/1538-4357/ad1690}

\bibitem[{{Balberg} {et~al.}(2013){Balberg}, {Sari}, \& {Loeb}}]{Balberg+13}
{Balberg}, S., {Sari}, R., \& {Loeb}, A. 2013, \mnras, 434, L26, \dodoi{10.1093/mnrasl/slt071}

\bibitem[{{Bar-Or} {et~al.}(2013){Bar-Or}, {Kupi}, \& {Alexander}}]{Bar-Or+13}
{Bar-Or}, B., {Kupi}, G., \& {Alexander}, T. 2013, \apj, 764, 52, \dodoi{10.1088/0004-637X/764/1/52}

\bibitem[{{Bartko} {et~al.}(2010){Bartko}, {Martins}, {Trippe}, {Fritz}, {Genzel}, {Ott}, {Eisenhauer}, {Gillessen}, {Paumard}, {Alexander}, {Dodds-Eden}, {Gerhard}, {Levin}, {Mascetti}, {Nayakshin}, {Perets}, {Perrin}, {Pfuhl}, {Reid}, {Rouan}, {Zilka}, \& {Sternberg}}]{Bartko+10}
{Bartko}, H., {Martins}, F., {Trippe}, S., {et~al.} 2010, \apj, 708, 834, \dodoi{10.1088/0004-637X/708/1/834}

\bibitem[{Behnel {et~al.}(2011)Behnel, Bradshaw, Citro, Dalcin, Seljebotn, \& Smith}]{cython:2011}
Behnel, S., Bradshaw, R., Citro, C., {et~al.} 2011, Computing in Science Engineering, 13, 31, \dodoi{10.1109/MCSE.2010.118}

\bibitem[{{Binney} \& {Tremaine}(2008)}]{BinneyTremaine}
{Binney}, J., \& {Tremaine}, S. 2008, {Galactic Dynamics: Second Edition}

\bibitem[{{Breivik} {et~al.}(2020{\natexlab{a}}){Breivik}, {Coughlin}, {Zevin}, {Rodriguez}, {Kremer}, {Ye}, {Andrews}, {Kurkowski}, {Digman}, {Larson}, \& {Rasio}}]{Breivik+20}
{Breivik}, K., {Coughlin}, S., {Zevin}, M., {et~al.} 2020{\natexlab{a}}, \apj, 898, 71, \dodoi{10.3847/1538-4357/ab9d85}

\bibitem[{{Breivik} {et~al.}(2020{\natexlab{b}}){Breivik}, {Coughlin}, {Zevin}, {Rodriguez}, {Kremer}, {Ye}, {Andrews}, {Kurkowski}, {Digman}, {Larson}, \& {Rasio}}]{Breivik2020}
---. 2020{\natexlab{b}}, \apj, 898, 71, \dodoi{10.3847/1538-4357/ab9d85}

\bibitem[{{Buchholz} {et~al.}(2009){Buchholz}, {Sch{\"o}del}, \& {Eckart}}]{Buchholz+09}
{Buchholz}, R.~M., {Sch{\"o}del}, R., \& {Eckart}, A. 2009, \aap, 499, 483, \dodoi{10.1051/0004-6361/200811497}

\bibitem[{Collette(2013)}]{collette_python_hdf5_2014}
Collette, A. 2013, Python and HDF5 (O'Reilly)

\bibitem[{Collette {et~al.}(2023)Collette, Kluyver, Caswell, Tocknell, Kieffer, Jelenak, Scopatz, Dale, Chen, VINCENT, Einhorn, payno, juliagarriga, Sciarelli, Valls, Ghosh, Pedersen, Kittisopikul, jakirkham, Raspaud, Danilevski, Abbasi, Readey, Mühlbauer, Paramonov, Chan, Schepper, Solé, jialin, \& Guest}]{h5py_7560547}
Collette, A., Kluyver, T., Caswell, T.~A., {et~al.} 2023, h5py/h5py: 3.8.0, 3.8.0,  Zenodo, \dodoi{10.5281/zenodo.7560547}

\bibitem[{Coughlin {et~al.}(2025)Coughlin, Breivik, Zevin, Wagg, Andrews, Rodriguez, Kimball, Poojan, Favero, mcdigman, Sharma, Perego, Ye, Martinez, Mandhai, elenagonzalez870, Chawla, MathieuVenet, \& 1nhtran}]{COSMIC_15855958}
Coughlin, S., Breivik, K., Zevin, M., {et~al.} 2025, COSMIC-PopSynth/COSMIC: v3.6.1, v3.6.1,  Zenodo, \dodoi{10.5281/zenodo.15855958}

\bibitem[{da~Costa-Luis {et~al.}(2024)da~Costa-Luis, Larroque, Altendorf, Mary, richardsheridan, Korobov, Yorav-Raphael, Ivanov, Bargull, Rodrigues, Shawn, Dektyarev, Górny, mjstevens777, Pagel, Zugnoni, JC, CrazyPython, Newey, Lee, pgajdos, Todd, Malmgren, redbug312, Desh, Nechaev, Boyle, Nordlund, MapleCCC, \& McCracken}]{tqdm_14231923}
da~Costa-Luis, C., Larroque, S.~K., Altendorf, K., {et~al.} 2024, tqdm: A fast, Extensible Progress Bar for Python and CLI, v4.67.1,  Zenodo, \dodoi{10.5281/zenodo.14231923}

\bibitem[{{Dale} \& {Davies}(2006)}]{DaleDavies}
{Dale}, J.~E., \& {Davies}, M.~B. 2006, \mnras, 366, 1424, \dodoi{10.1111/j.1365-2966.2005.09937.x}

\bibitem[{{Dale} {et~al.}(2009){Dale}, {Davies}, {Church}, \& {Freitag}}]{Dale+09}
{Dale}, J.~E., {Davies}, M.~B., {Church}, R.~P., \& {Freitag}, M. 2009, \mnras, 393, 1016, \dodoi{10.1111/j.1365-2966.2008.14254.x}

\bibitem[{{Davies} {et~al.}(1998){Davies}, {Blackwell}, {Bailey}, \& {Sigurdsson}}]{Davies+98}
{Davies}, M.~B., {Blackwell}, R., {Bailey}, V.~C., \& {Sigurdsson}, S. 1998, \mnras, 301, 745, \dodoi{10.1046/j.1365-8711.1998.02027.x}

\bibitem[{{Davies} {et~al.}(2011){Davies}, {Church}, {Malmberg}, {Nzoke}, {Dale}, \& {Freitag}}]{Davies+11}
{Davies}, M.~B., {Church}, R.~P., {Malmberg}, D., {et~al.} 2011, in Astronomical Society of the Pacific Conference Series, Vol. 439, The Galactic Center: a Window to the Nuclear Environment of Disk Galaxies, ed. M.~R. {Morris}, Q.~D. {Wang}, \& F.~{Yuan}, 212.
\newblock \doarXiv{1002.0338}

\bibitem[{{Do} {et~al.}(2009){Do}, {Ghez}, {Morris}, {Lu}, {Matthews}, {Yelda}, \& {Larkin}}]{Do+09}
{Do}, T., {Ghez}, A.~M., {Morris}, M.~R., {et~al.} 2009, \apj, 703, 1323, \dodoi{10.1088/0004-637X/703/2/1323}

\bibitem[{{Duncan} \& {Shapiro}(1983)}]{DuncanShapiro83}
{Duncan}, M.~J., \& {Shapiro}, S.~L. 1983, \apj, 268, 565, \dodoi{10.1086/160980}

\bibitem[{{Ferrarese} \& {Ford}(2005)}]{FerrareseFord05}
{Ferrarese}, L., \& {Ford}, H. 2005, \ssr, 116, 523, \dodoi{10.1007/s11214-005-3947-6}

\bibitem[{{Freitag}(2008)}]{Freitag+08}
{Freitag}, M. 2008, in Astronomical Society of the Pacific Conference Series, Vol. 387, Massive Star Formation: Observations Confront Theory, ed. H.~{Beuther}, H.~{Linz}, \& T.~{Henning}, 247, \dodoi{10.48550/arXiv.0711.4057}

\bibitem[{{Freitag} {et~al.}(2006){Freitag}, {Amaro-Seoane}, \& {Kalogera}}]{Freitag+06}
{Freitag}, M., {Amaro-Seoane}, P., \& {Kalogera}, V. 2006, \apj, 649, 91, \dodoi{10.1086/506193}

\bibitem[{{Freitag} \& {Benz}(2002{\natexlab{a}})}]{Freitag+02}
{Freitag}, M., \& {Benz}, W. 2002{\natexlab{a}}, \aap, 394, 345, \dodoi{10.1051/0004-6361:20021142}

\bibitem[{{Freitag} \& {Benz}(2002{\natexlab{b}})}]{FreitagBenz_confproceedings_2002}
{Freitag}, M., \& {Benz}, W. 2002{\natexlab{b}}, in Astronomical Society of the Pacific Conference Series, Vol. 263, Stellar Collisions, Mergers and their Consequences, ed. M.~M. {Shara}, 261, \dodoi{10.48550/arXiv.astro-ph/0101186}

\bibitem[{{Freitag} \& {Benz}(2005)}]{FreitagBenz}
---. 2005, \mnras, 358, 1133, \dodoi{10.1111/j.1365-2966.2005.08770.x}

\bibitem[{{Freitag} {et~al.}(2008){Freitag}, {Dale}, {Church}, \& {Davies}}]{Freitag+08_confproceeding}
{Freitag}, M., {Dale}, J.~E., {Church}, R.~P., \& {Davies}, M.~B. 2008, in IAU Symposium, Vol. 245, Formation and Evolution of Galaxy Bulges, ed. M.~{Bureau}, E.~{Athanassoula}, \& B.~{Barbuy}, 211--214, \dodoi{10.1017/S1743921308017675}

\bibitem[{{Gallego-Cano} {et~al.}(2018){Gallego-Cano}, {Sch{\"o}del}, {Dong}, {Nogueras-Lara}, {Gallego-Calvente}, {Amaro-Seoane}, \& {Baumgardt}}]{Gallego-Cano+18}
{Gallego-Cano}, E., {Sch{\"o}del}, R., {Dong}, H., {et~al.} 2018, \aap, 609, A26, \dodoi{10.1051/0004-6361/201730451}

\bibitem[{{Gallego-Cano} {et~al.}(2020){Gallego-Cano}, {Sch{\"o}del}, {Nogueras-Lara}, {Dong}, {Shahzamanian}, {Fritz}, {Gallego-Calvente}, \& {Neumayer}}]{Gallego+20}
{Gallego-Cano}, E., {Sch{\"o}del}, R., {Nogueras-Lara}, F., {et~al.} 2020, \aap, 634, A71, \dodoi{10.1051/0004-6361/201935303}

\bibitem[{{Genzel} {et~al.}(2010){Genzel}, {Eisenhauer}, \& {Gillessen}}]{Genzel+10}
{Genzel}, R., {Eisenhauer}, F., \& {Gillessen}, S. 2010, Reviews of Modern Physics, 82, 3121, \dodoi{10.1103/RevModPhys.82.3121}

\bibitem[{{Genzel} {et~al.}(2003){Genzel}, {Sch{\"o}del}, {Ott}, {Eisenhauer}, {Hofmann}, {Lehnert}, {Eckart}, {Alexander}, {Sternberg}, {Lenzen}, {Cl{\'e}net}, {Lacombe}, {Rouan}, {Renzini}, \& {Tacconi-Garman}}]{Genzel+03}
{Genzel}, R., {Sch{\"o}del}, R., {Ott}, T., {et~al.} 2003, \apj, 594, 812, \dodoi{10.1086/377127}

\bibitem[{{Ghez} {et~al.}(2005){Ghez}, {Salim}, {Hornstein}, {Tanner}, {Lu}, {Morris}, {Becklin}, \& {Duch{\^e}ne}}]{Ghez+05}
{Ghez}, A.~M., {Salim}, S., {Hornstein}, S.~D., {et~al.} 2005, \apj, 620, 744, \dodoi{10.1086/427175}

\bibitem[{{Ghez} {et~al.}(2003){Ghez}, {Duch{\^e}ne}, {Matthews}, {Hornstein}, {Tanner}, {Larkin}, {Morris}, {Becklin}, {Salim}, {Kremenek}, {Thompson}, {Soifer}, {Neugebauer}, \& {McLean}}]{Ghez+03}
{Ghez}, A.~M., {Duch{\^e}ne}, G., {Matthews}, K., {et~al.} 2003, \apjl, 586, L127, \dodoi{10.1086/374804}

\bibitem[{{Ghez} {et~al.}(2008){Ghez}, {Salim}, {Weinberg}, {Lu}, {Do}, {Dunn}, {Matthews}, {Morris}, {Yelda}, {Becklin}, {Kremenek}, {Milosavljevic}, \& {Naiman}}]{Ghez+08}
{Ghez}, A.~M., {Salim}, S., {Weinberg}, N.~N., {et~al.} 2008, \apj, 689, 1044, \dodoi{10.1086/592738}

\bibitem[{{Gibson} {et~al.}(2024){Gibson}, {K{\i}ro{\u{g}}lu}, {Lombardi}, {Rose}, {Vanderzyden}, {Mockler}, {Gallegos-Garcia}, {Kremer}, {Ramirez-Ruiz}, \& {Rasio}}]{Gibson+24}
{Gibson}, C., {K{\i}ro{\u{g}}lu}, F., {Lombardi}, J.~C., J., {et~al.} 2024, arXiv e-prints, arXiv:2410.02146, \dodoi{10.48550/arXiv.2410.02146}

\bibitem[{{Gillessen} {et~al.}(2009){Gillessen}, {Eisenhauer}, {Trippe}, {Alexand er}, {Genzel}, {Martins}, \& {Ott}}]{Gillessen+09}
{Gillessen}, S., {Eisenhauer}, F., {Trippe}, S., {et~al.} 2009, \apj, 692, 1075, \dodoi{10.1088/0004-637X/692/2/1075}

\bibitem[{{Gillessen} {et~al.}(2017){Gillessen}, {Plewa}, {Eisenhauer}, {Sari}, {Waisberg}, {Habibi}, {Pfuhl}, {George}, {Dexter}, {von Fellenberg}, {Ott}, \& {Genzel}}]{Gillessen+17}
{Gillessen}, S., {Plewa}, P.~M., {Eisenhauer}, F., {et~al.} 2017, \apj, 837, 30, \dodoi{10.3847/1538-4357/aa5c41}

\bibitem[{Glebbeek {et~al.}(2013)Glebbeek, Gaburov, Portegies~Zwart, \& Pols}]{Glebbeek+13}
Glebbeek, E., Gaburov, E., Portegies~Zwart, S., \& Pols, O.~R. 2013, Monthly Notices of the Royal Astronomical Society, 434, 3497, \dodoi{10.1093/mnras/stt1268}

\bibitem[{Gommers {et~al.}(2025)Gommers, Virtanen, Haberland, Burovski, Reddy, Weckesser, Oliphant, Nelson, Cournapeau, alexbrc, Roy, Polat, Peterson, Wilson, Colley, endolith, Mayorov, van~der Walt, Bowhay, Brett, Laxalde, Steppi, Larson, Sakai, Millman, Lars, peterbell10, Carey, van Mulbregt, \& eric jones}]{scipy_17467817}
Gommers, R., Virtanen, P., Haberland, M., {et~al.} 2025, scipy/scipy: SciPy 1.16.3, v1.16.3,  Zenodo, \dodoi{10.5281/zenodo.17467817}

\bibitem[{{Guillochon} \& {Ramirez-Ruiz}(2013)}]{Guillochon+13}
{Guillochon}, J., \& {Ramirez-Ruiz}, E. 2013, \apj, 767, 25, \dodoi{10.1088/0004-637X/767/1/25}

\bibitem[{{Guillochon} {et~al.}(2009){Guillochon}, {Ramirez-Ruiz}, {Rosswog}, \& {Kasen}}]{Guillochon+09}
{Guillochon}, J., {Ramirez-Ruiz}, E., {Rosswog}, S., \& {Kasen}, D. 2009, \apj, 705, 844, \dodoi{10.1088/0004-637X/705/1/844}

\bibitem[{{G{\"u}rkan} {et~al.}(2006){G{\"u}rkan}, {Fregeau}, \& {Rasio}}]{Gurkan+06}
{G{\"u}rkan}, M.~A., {Fregeau}, J.~M., \& {Rasio}, F.~A. 2006, \apjl, 640, L39, \dodoi{10.1086/503295}

\bibitem[{{Habibi} {et~al.}(2017){Habibi}, {Gillessen}, {Martins}, {Eisenhauer}, {Plewa}, {Pfuhl}, {George}, {Dexter}, {Waisberg}, {Ott}, {von Fellenberg}, {Baub{\"o}ck}, {Jimenez-Rosales}, \& {Genzel}}]{Habibi+17}
{Habibi}, M., {Gillessen}, S., {Martins}, F., {et~al.} 2017, \apj, 847, 120, \dodoi{10.3847/1538-4357/aa876f}

\bibitem[{Harris {et~al.}(2020)Harris, Millman, van~der Walt, Gommers, Virtanen, Cournapeau, Wieser, Taylor, Berg, Smith, Kern, Picus, Hoyer, van Kerkwijk, Brett, Haldane, del R{\'{i}}o, Wiebe, Peterson, G{\'{e}}rard-Marchant, Sheppard, Reddy, Weckesser, Abbasi, Gohlke, \& Oliphant}]{numpy}
Harris, C.~R., Millman, K.~J., van~der Walt, S.~J., {et~al.} 2020, Nature, 585, 357, \dodoi{10.1038/s41586-020-2649-2}

\bibitem[{{Hills}(1975)}]{Hills1975}
{Hills}, J.~G. 1975, \nat, 254, 295, \dodoi{10.1038/254295a0}

\bibitem[{{Hopman} \& {Alexander}(2005)}]{HopmanAlexander05}
{Hopman}, C., \& {Alexander}, T. 2005, \apj, 629, 362, \dodoi{10.1086/431475}

\bibitem[{Hunter(2007)}]{Hunter:2007}
Hunter, J.~D. 2007, Computing in Science \& Engineering, 9, 90, \dodoi{10.1109/MCSE.2007.55}

\bibitem[{{Keshet} {et~al.}(2009){Keshet}, {Hopman}, \& {Alexander}}]{Keshet+09}
{Keshet}, U., {Hopman}, C., \& {Alexander}, T. 2009, \apjl, 698, L64, \dodoi{10.1088/0004-637X/698/1/L64}

\bibitem[{{Kormendy}(2004)}]{Kormendy04}
{Kormendy}, J. 2004, in Coevolution of Black Holes and Galaxies, ed. L.~C. {Ho}, 1.
\newblock \doarXiv{astro-ph/0306353}

\bibitem[{{Kormendy} \& {Ho}(2013)}]{KormendyHo13}
{Kormendy}, J., \& {Ho}, L.~C. 2013, \araa, 51, 511, \dodoi{10.1146/annurev-astro-082708-101811}

\bibitem[{Kushnir {et~al.}(2013)Kushnir, Katz, Dong, Livne, \& Fernández}]{Kushnir+13}
Kushnir, D., Katz, B., Dong, S., Livne, E., \& Fernández, R. 2013, The Astrophysical Journal, 778, L37, \dodoi{10.1088/2041-8205/778/2/l37}

\bibitem[{{Lai} {et~al.}(1993){Lai}, {Rasio}, \& {Shapiro}}]{Lai+93}
{Lai}, D., {Rasio}, F.~A., \& {Shapiro}, S.~L. 1993, \apj, 412, 593, \dodoi{10.1086/172946}

\bibitem[{{Levin} \& {Beloborodov}(2003)}]{Levin+03}
{Levin}, Y., \& {Beloborodov}, A.~M. 2003, \apjl, 590, L33, \dodoi{10.1086/376675}

\bibitem[{{Linial} \& {Sari}(2022)}]{LinialSari22}
{Linial}, I., \& {Sari}, R. 2022, \apj, 940, 101, \dodoi{10.3847/1538-4357/ac9bfd}

\bibitem[{{Lombardi} {et~al.}(2002){Lombardi}, {Warren}, {Rasio}, {Sills}, \& {Warren}}]{Lombardi+02}
{Lombardi}, James~C., J., {Warren}, J.~S., {Rasio}, F.~A., {Sills}, A., \& {Warren}, A.~R. 2002, \apj, 568, 939, \dodoi{10.1086/339060}

\bibitem[{{Lombardi} {et~al.}(1996){Lombardi}, {Rasio}, \& {Shapiro}}]{Lombardi+96}
{Lombardi}, Jr., J.~C., {Rasio}, F.~A., \& {Shapiro}, S.~L. 1996, \apj, 468, 797, \dodoi{10.1086/177736}

\bibitem[{MacLeod {et~al.}(2013)MacLeod, Ramirez-Ruiz, Grady, \& Guillochon}]{macleod_spoon-feeding_2013}
MacLeod, M., Ramirez-Ruiz, E., Grady, S., \& Guillochon, J. 2013, The Astrophysical Journal, 777, 133, \dodoi{10.1088/0004-637X/777/2/133}

\bibitem[{Marín-Franch {et~al.}(2009)Marín-Franch, Aparicio, Piotto, Rosenberg, Chaboyer, Sarajedini, Siegel, Anderson, Bedin, Dotter, Hempel, King, Majewski, Milone, Paust, \& Reid}]{MarinFranch+09}
Marín-Franch, A., Aparicio, A., Piotto, G., {et~al.} 2009, The Astrophysical Journal, 694, 1498–1516, \dodoi{10.1088/0004-637x/694/2/1498}

\bibitem[{{Mastrobuono-Battisti} {et~al.}(2021){Mastrobuono-Battisti}, {Church}, \& {Davies}}]{Mastrobuono-B}
{Mastrobuono-Battisti}, A., {Church}, R.~P., \& {Davies}, M.~B. 2021, \mnras, 505, 3314, \dodoi{10.1093/mnras/stab1409}

\bibitem[{{Merritt}(2013)}]{Merritt2013}
{Merritt}, D. 2013, {Dynamics and Evolution of Galactic Nuclei} (Princeton University Press)

\bibitem[{{Naoz} {et~al.}(2022){Naoz}, {Rose}, {Michaely}, {Melchor}, {Ramirez-Ruiz}, {Mockler}, \& {Schnittman}}]{Naoz+22}
{Naoz}, S., {Rose}, S.~C., {Michaely}, E., {et~al.} 2022, \apjl, 927, L18, \dodoi{10.3847/2041-8213/ac574b}

\bibitem[{pandas~development team(2025)}]{pandas_17806077}
pandas~development team, T. 2025, pandas-dev/pandas: Pandas, v3.0.0rc0,  Zenodo, \dodoi{10.5281/zenodo.17806077}

\bibitem[{{Portegies Zwart} \& {McMillan}(2000)}]{PortegiesZwart}
{Portegies Zwart}, S.~F., \& {McMillan}, S. L.~W. 2000, \apjl, 528, L17, \dodoi{10.1086/312422}

\bibitem[{Price-Whelan \& Foreman-Mackey(2017)}]{schwimmbad}
Price-Whelan, A.~M., \& Foreman-Mackey, D. 2017, The Journal of Open Source Software, 2, \dodoi{10.21105/joss.00357}

\bibitem[{{Raskin} {et~al.}(2009){Raskin}, {Timmes}, {Scannapieco}, {Diehl}, \& {Fryer}}]{Raskin+09}
{Raskin}, C., {Timmes}, F.~X., {Scannapieco}, E., {Diehl}, S., \& {Fryer}, C. 2009, \mnras, 399, L156, \dodoi{10.1111/j.1745-3933.2009.00743.x}

\bibitem[{{Rauch}(1999)}]{Rauch99}
{Rauch}, K.~P. 1999, \apj, 514, 725, \dodoi{10.1086/306953}

\bibitem[{{Rees}(1988)}]{Rees1988}
{Rees}, M.~J. 1988, \nat, 333, 523, \dodoi{10.1038/333523a0}

\bibitem[{{Rose} {et~al.}(2025){Rose}, {Lombardi}, {Gonz{\'a}lez Prieto}, {K{\i}ro{\u{g}}lu}, \& {Rasio}}]{Rose+25}
{Rose}, S.~C., {Lombardi}, Jr., J.~C., {Gonz{\'a}lez Prieto}, E., {K{\i}ro{\u{g}}lu}, F., \& {Rasio}, F.~A. 2025, arXiv e-prints, arXiv:2511.01811, \dodoi{10.48550/arXiv.2511.01811}

\bibitem[{{Rose} \& {MacLeod}(2024)}]{RoseMacLeod24}
{Rose}, S.~C., \& {MacLeod}, M. 2024, \apjl, 963, L17, \dodoi{10.3847/2041-8213/ad251f}

\bibitem[{{Rose} \& {Mockler}(2025)}]{RoseMockler25}
{Rose}, S.~C., \& {Mockler}, B. 2025, \apjl, 985, L40, \dodoi{10.3847/2041-8213/add266}

\bibitem[{{Rose} {et~al.}(2020){Rose}, {Naoz}, {Gautam}, {Ghez}, {Do}, {Chu}, \& {Becklin}}]{Rose+20}
{Rose}, S.~C., {Naoz}, S., {Gautam}, A.~K., {et~al.} 2020, \apj, 904, 113, \dodoi{10.3847/1538-4357/abc557}

\bibitem[{{Rose} {et~al.}(2022){Rose}, {Naoz}, {Sari}, \& {Linial}}]{Rose+22}
{Rose}, S.~C., {Naoz}, S., {Sari}, R., \& {Linial}, I. 2022, \apjl, 929, L22, \dodoi{10.3847/2041-8213/ac6426}

\bibitem[{{Rose} {et~al.}(2023){Rose}, {Naoz}, {Sari}, \& {Linial}}]{Rose+23}
---. 2023, arXiv e-prints, arXiv:2304.10569, \dodoi{10.48550/arXiv.2304.10569}

\bibitem[{Rosswog {et~al.}(2009)Rosswog, Ramirez-Ruiz, \& Hix}]{rosswog_tidal_2009}
Rosswog, S., Ramirez-Ruiz, E., \& Hix, W.~R. 2009, The Astrophysical Journal, 695, 404, \dodoi{10.1088/0004-637X/695/1/404}

\bibitem[{{Rubin} \& {Loeb}(2011)}]{RubinLoeb}
{Rubin}, D., \& {Loeb}, A. 2011, Advances in Astronomy, 2011, 174105, \dodoi{10.1155/2011/174105}

\bibitem[{{Ryu} {et~al.}(2024){Ryu}, {Amaro Seoane}, {Taylor}, \& {Ohlmann}}]{Ryu+24b}
{Ryu}, T., {Amaro Seoane}, P., {Taylor}, A.~M., \& {Ohlmann}, S.~T. 2024, \mnras, 528, 6193, \dodoi{10.1093/mnras/stae396}

\bibitem[{{Sch{\"o}del} {et~al.}(2014){Sch{\"o}del}, {Feldmeier}, {Kunneriath}, {Stolovy}, {Neumayer}, {Amaro-Seoane}, \& {Nishiyama}}]{Schodel+14}
{Sch{\"o}del}, R., {Feldmeier}, A., {Kunneriath}, D., {et~al.} 2014, \aap, 566, A47, \dodoi{10.1051/0004-6361/201423481}

\bibitem[{{Sch{\"o}del} {et~al.}(2018){Sch{\"o}del}, {Gallego-Cano}, {Dong}, {Nogueras-Lara}, {Gallego-Calvente}, {Amaro-Seoane}, \& {Baumgardt}}]{Schodel+18}
{Sch{\"o}del}, R., {Gallego-Cano}, E., {Dong}, H., {et~al.} 2018, \aap, 609, A27, \dodoi{10.1051/0004-6361/201730452}

\bibitem[{{Sch{\"o}del} {et~al.}(2003){Sch{\"o}del}, {Genzel}, {Ott}, \& {Eckart}}]{Schodel+03}
{Sch{\"o}del}, R., {Genzel}, R., {Ott}, T., \& {Eckart}, A. 2003, Astronomische Nachrichten Supplement, 324, 535, \dodoi{10.1002/asna.200385048}

\bibitem[{{Sch{\"o}del} {et~al.}(2020){Sch{\"o}del}, {Nogueras-Lara}, {Gallego-Cano}, {Shahzamanian}, {Gallego-Calvente}, \& {Gardini}}]{Schodel+20}
{Sch{\"o}del}, R., {Nogueras-Lara}, F., {Gallego-Cano}, E., {et~al.} 2020, \aap, 641, A102, \dodoi{10.1051/0004-6361/201936688}

\bibitem[{{Sills} {et~al.}(2001){Sills}, {Faber}, {Lombardi}, {Rasio}, \& {Warren}}]{Sills+01}
{Sills}, A., {Faber}, J.~A., {Lombardi}, James~C., J., {Rasio}, F.~A., \& {Warren}, A.~R. 2001, \apj, 548, 323, \dodoi{10.1086/318689}

\bibitem[{Sills {et~al.}(2010)Sills, Kologera, \& van~der Sluys}]{Sills+10}
Sills, A., Kologera, V., \& van~der Sluys, M. 2010, in AIP Conference Proceedings (AIP), 105–112, \dodoi{10.1063/1.3536351}

\bibitem[{{Sills} {et~al.}(1997){Sills}, {Lombardi}, {Bailyn}, {Demarque}, {Rasio}, \& {Shapiro}}]{Sills+97}
{Sills}, A., {Lombardi}, James~C., J., {Bailyn}, C.~D., {et~al.} 1997, \apj, 487, 290, \dodoi{10.1086/304588}

\bibitem[{{Spitzer}(1987)}]{Spitzer1987}
{Spitzer}, L. 1987, {Dynamical evolution of globular clusters}

\bibitem[{{Stone} {et~al.}(2017){Stone}, {K{\"u}pper}, \& {Ostriker}}]{Stone+17}
{Stone}, N.~C., {K{\"u}pper}, A. H.~W., \& {Ostriker}, J.~P. 2017, \mnras, 467, 4180, \dodoi{10.1093/mnras/stx097}

\bibitem[{{Tremaine} {et~al.}(2002){Tremaine}, {Gebhardt}, {Bender}, {Bower}, {Dressler}, {Faber}, {Filippenko}, {Green}, {Grillmair}, {Ho}, {Kormendy}, {Lauer}, {Magorrian}, {Pinkney}, \& {Richstone}}]{Tremaine+02}
{Tremaine}, S., {Gebhardt}, K., {Bender}, R., {et~al.} 2002, \apj, 574, 740, \dodoi{10.1086/341002}

\bibitem[{Van~Rossum \& Drake(2009)}]{python}
Van~Rossum, G., \& Drake, F.~L. 2009, Python 3 Reference Manual (Scotts Valley, CA: CreateSpace)

\bibitem[{Virtanen {et~al.}(2020)Virtanen, Gommers, Oliphant, Haberland, Reddy, Cournapeau, Burovski, Peterson, Weckesser, Bright, {van der Walt}, Brett, Wilson, Millman, Mayorov, Nelson, Jones, Kern, Larson, Carey, Polat, Feng, Moore, {VanderPlas}, Laxalde, Perktold, Cimrman, Henriksen, Quintero, Harris, Archibald, Ribeiro, Pedregosa, {van Mulbregt}, \& {SciPy 1.0 Contributors}}]{2020SciPy-NMeth}
Virtanen, P., Gommers, R., Oliphant, T.~E., {et~al.} 2020, Nature Methods, 17, 261, \dodoi{10.1038/s41592-019-0686-2}

\bibitem[{Wagg {et~al.}(2025)Wagg, Broekgaarden, Van-Lane, Wu, \& Gültekin}]{software-citation-station-zenodo}
Wagg, T., Broekgaarden, F., Van-Lane, P., Wu, K., \& Gültekin, K. 2025, TomWagg/software-citation-station: v1.4, v1.4,  Zenodo, \dodoi{10.5281/zenodo.17654855}

\bibitem[{{Wagg} \& {Broekgaarden}(2024)}]{software-citation-station-paper}
{Wagg}, T., \& {Broekgaarden}, F.~S. 2024, arXiv e-prints, arXiv:2406.04405.
\newblock \doarXiv{2406.04405}

\bibitem[{Waskom(2021)}]{Waskom2021}
Waskom, M.~L. 2021, Journal of Open Source Software, 6, 3021, \dodoi{10.21105/joss.03021}

\bibitem[{{W}es {M}c{K}inney(2010)}]{mckinney-proc-scipy-2010}
{W}es {M}c{K}inney. 2010, in {P}roceedings of the 9th {P}ython in {S}cience {C}onference, ed. {S}t\'efan van~der {W}alt \& {J}arrod {M}illman, 56 -- 61, \dodoi{10.25080/Majora-92bf1922-00a}

\end{thebibliography}
\bibliographystyle{aasjournal}

\end{document}